\documentclass[sigconf]{acmart}
\usepackage{subcaption}
\usepackage[british]{babel}

\AtBeginDocument{%
  \providecommand\BibTeX{{%
    \normalfont B\kern-0.5em{\scshape i\kern-0.25em b}\kern-0.8em\TeX}}}

\setcopyright{iw3c2w3}

\begin{document}

\title{Information Extraction From Co-Occurring Similar Entities}

\author{Nicolas Heist}
\email{nico@informatik.uni-mannheim.de}
\orcid{0000-0002-4354-9138}
\affiliation{%
  \institution{Data and Web Science Group}
  \city{University of Mannheim}
  \country{Germany}
}

\author{Heiko Paulheim}
\email{heiko@informatik.uni-mannheim.de}
\orcid{0000-0003-4386-8195}
\affiliation{%
  \institution{Data and Web Science Group}
  \city{University of Mannheim}
  \country{Germany}
}

\renewcommand{\shortauthors}{N. Heist and H. Paulheim}
\begin{abstract}
Knowledge about entities and their interrelations is a crucial factor of success for tasks like question answering or text summarization. Publicly available knowledge graphs like Wikidata or DBpedia are, however, far from being complete. In this paper, we explore how information extracted from similar entities that co-occur in structures like tables or lists can help to increase the coverage of such knowledge graphs. In contrast to existing approaches, we do not focus on relationships within a listing (e.g., between two entities in a table row) but on the relationship between a listing's subject entities and the context of the listing. To that end, we propose a descriptive rule mining approach that uses distant supervision to derive rules for these relationships based on a listing's context. Extracted from a suitable data corpus, the rules can be used to extend a knowledge graph with novel entities and assertions. In our experiments we demonstrate that the approach is able to extract up to 3M novel entities and 30M additional assertions from listings in Wikipedia. We find that the extracted information is of high quality and thus suitable to extend Wikipedia-based knowledge graphs like DBpedia, YAGO, and CaLiGraph. For the case of DBpedia, this would result in an increase of covered entities by roughly 50\%.
\end{abstract}


\copyrightyear{2021}
\acmYear{2021}
\acmConference[WWW '21]{Proceedings of the Web Conference 2021}{April 19--23, 2021}{Ljubljana, Slovenia}
\acmBooktitle{Proceedings of the Web Conference 2021 (WWW '21), April 19--23, 2021, Ljubljana, Slovenia}
\acmPrice{}
\acmDOI{10.1145/3442381.3449836}
\acmISBN{978-1-4503-8312-7/21/04}


\maketitle

\section{Introduction}
\subsection{Motivation and Problem}
\label{subsec:motivation-and-problem}
In tasks like question answering, text summarization, or entity disambiguation, it is essential to have background information about the involved entities. With entity linking tools like DBpedia Spotlight \cite{mendes2011dbpedia} or Falcon \cite{sakor2020falcon}, one can easily identify named entities in text and retrieve the respective entity in a background entity hub of the linking tool (e.g. in a wiki like Wikipedia or in a knowledge graph like DBpedia \cite{lehmann2015dbpedia}). This is, however, only possible if the entity in question is contained in the respective entity hub \cite{van2016evaluating}.

The trend of entities added to publicly available knowledge graphs in recent years indicates that they are far from being complete. The number of entities in Wikidata \cite{vrandevcic2014wikidata}, for example, grew by 37\% in the time from October 2019 (61.7M) to October 2020 (84.5M). In the same time, the number of statements increased by 41\% from 770M to 1085M.\footnote{\url{https://tools.wmflabs.org/wikidata-todo/stats.php}}  According to \cite{heist2020knowledge}, Wikidata describes the largest number of entities and comprises -- in terms of entities -- other open knowledge graphs to a large extent. Consequently, this problem applies to all public knowledge graphs, and particularly so for long-tail and emerging entities \cite{farber2016emerging}.

Automatic information extraction approaches can help mitigating this problem if the approaches can make sure that the extracted information is of high quality. While the performance of open information extraction systems (i.e. systems that extract information from general web text) has improved in recent years \cite{liu2020extracting,stanovsky2018supervised,del2013clausie}, the quality of extracted information has not yet reached a level where an integration into knowledge graphs like DBpedia should be done without further filtering.

\begin{figure}[t]
  \centering
  \includegraphics[width=.5\linewidth]{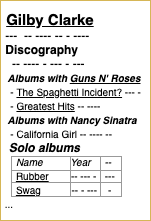}
  \caption{Simplified view on the Wikipedia page of Gilby Clarke with a focus on its title, sections, and listings.}
  \label{fig:running-example}
  \Description{A simplified view on the Wikipedia page of Gilby Clarke is shown which contains only the page title, sections, and listings of the page. The focus is set on the section "Discography" with its subsections "Albums with Guns N' Roses", "Albums with Nancy Sinatra", and "Solo albums". Each of the subsections contain lists or tables with instances of albums.}
\end{figure}

The extraction of information from semi-structured data is in general less error-prone and already proved to yield high-quality results as, for example, DBpedia itself is extracted primarily from Wikipedia infoboxes; further approaches use the category system of Wikipedia \cite{suchanek2007yago,heist2019uncovering,xu2016learning} or its list pages \cite{paulheim2013extending,heist2020entity}. Many more approaches focus on tables (in Wikipedia or the web) as semi-structured data source to extract entities and relations (see \cite{zhang2020web} for a comprehensive survey). The focus of recent web table-based approaches like Zhang et al. \cite{zhang2020novel} is set on recognizing entities and relationships within a table. Considering Fig.~\ref{fig:running-example}, the table below the section \textit{Solo albums} may be used to discover the publication years of albums (relation extraction) or discover additional unknown albums that are listed in further rows below \textit{Rubber} and \textit{Swag} (entity and type detection).

The focus of this paper is broader with respect to two dimensions: First, we extract information from any kind of structure where similar entities co-occur. In Fig.~\ref{fig:running-example}, we would consider both tables and lists (e.g. the list in the section \textit{Albums with Guns N' Roses}). We refer to these co-occurrence structures as listings. Second, we consider only the subject entities (SE) of listings. In our previous work we defined SE with respect to Wikipedia list pages as \textit{"the instances of the concept expressed by the list page"} \cite{heist2020entity}. Considering the \textit{List of Japanese speculative fiction writers}, its SE comprise all Japanese speculative fiction writers mentioned in listings of the page. While in \cite{heist2020entity} the concept of SE is made explicit by the list page, we deal with arbitrary listings in this paper. We thus assume the concept may not be explicit or it may be indicated as part of the page in which the listing appears (e.g. in the table header, or the page title). Therefore, to each entity in a listing appearing as instance to a common concept, we will further refer as subject entity. The purpose of this work is to exploit the relationship between the SE of a listing and the listing context. For Fig.~\ref{fig:running-example}, this means we extract that all SE on the page's listings are albums with the artist \textit{Gilby Clarke}, that \textit{The Spaghetti Incident?} is an album by \textit{Guns N' Roses}, and so on.

To that end, we propose to learn these characteristics of a listing with respect to the types and contextual relations of its SE. In an ideal setting we know the SE of a listing and we are able to retrieve all information about them from a knowledge graph -- the characteristics of a listing are then simply the types and relations that are shared by all SE. But uncertainty is introduced by several factors:
\begin{itemize}
    \item SE can only be determined heuristically. In previous work \cite{heist2020entity}, we achieved a precision of 90\% for the recognition of SE in Wikipedia listings.
    \item Cross-domain knowledge graphs are not complete. According to the open world assumption (OWA), the absence of a fact in a knowledge graph does not imply its incorrectness.
    \item Web tables have a median of 6 rows,\footnote{According to the WDC Web Table Corpus 2015: \url{http://webdatacommons.org/webtables/}.} and Wikipedia listings have a median of 8 rows. Consequently, many listings only have a small number of SE from which the characteristics can be inferred.
\end{itemize}

As a result, considering each listing in isolation either leads to a substantial loss of information (as listings with insufficient background information are disregarded) or to a high generalization error (as decisions are made based on insufficient background information).

We observe that the context of a listing is often a strong indicator for its characteristics. In Fig.~\ref{fig:running-example}, the title of the top section \textit{Discography} indicates that its listings contain some kind of musical works, and the section title \textit{Albums with Guns N' Roses} provides more detailed information. Our second observation is that these patterns repeat when looking at a coherent data corpus. The Wikipedia page of \textit{Axl Rose},\footnote{\url{https://en.wikipedia.org/wiki/Axl_Rose}} for example, contains the same constellation of sections.

Considering listing characteristics with respect to their context can thus yield in more general insights than considering every listing in isolation. For example, the musical works of many artists in Wikipedia are listed under the top section \textit{Discography}. Hence, we could learn the axioms
\begin{equation}\label{eq:axiom-example-1}
\exists topSection.\{\textrm{"Discography"}\} \sqsubseteq \mathtt{MusicalWork}
\end{equation}
and
\begin{equation}\label{eq:axiom-example-2}
\exists topSection.\{\textrm{"Discography"}\} \sqsubseteq \exists artist.\{\textrm{<}PageEntity\textrm{>}\}
\end{equation}
which are then applicable to any listing with the top section \textit{Discography} in Wikipedia.

\subsection{Approach and Contributions}
In this work, we frame the task of finding descriptive rules for listings based on their context as association rule mining problem \cite{agrawal1993mining}. We define rule metrics that take the inherent uncertainty into account and make sure that rules are frequent (rule support), correct (rule confidence), and consistent over all listings (rule consistency). Furthermore, we present an approach that executes the complete pipeline from identification of SE to the extraction of novel entities and assertions with Wikipedia as data corpus. To find a reasonable balance between correctness and coverage of the rules, we set the thresholds based on a heuristic that takes the distribution of named entity tags over entities as well as existing knowledge in a knowledge graph into account. Applying the approach, we show that we can enhance the knowledge graphs DBpedia with up to 2.9M entities and 8.3M assertions, and CaLiGraph\footnote{\url{http://caligraph.org}} with up to 3M entities and 30.4M assertions with an overall correctness of more than 90\%.

To summarize, the contributions of this paper are as follows:
\begin{itemize}
    \item We formulate the task of information extraction from co-occurring similar entities in listings and show how to derive descriptive rules for listing characteristics based on the listing context (Sec.~\ref{sec:ie-from-cooccurrences}).
    \item We present an approach that learns descriptive rules for listings in Wikipedia and is capable of extracting several millions of novel entities and assertions for Wikipedia-based knowledge graphs (Sec.~\ref{sec:exploiting-co-occurrences-in-Wikipedia}).
    \item In our evaluation we demonstrate the high quality of the extracted information and analyze the shortcomings of the approach (Sec.~\ref{sec:evaluation}).
\end{itemize}

The produced code is part of the CaLiGraph extraction framework and publicly available.\footnote{\url{https://github.com/nheist/CaLiGraph}}

\section{Related Work}
The work presented in this paper is a flavour of \emph{knowledge graph completion}, more precisely, of adding new entities to a knowledge graph \cite{paulheim2017knowledge}. We use rules based on page context to infer facts about co-occurring entities. In particular, we focus on co-occurrence of entities within document listings, where co-occurrence refers to proximity in page layout. Hence, in this section, we discuss related works w.r.t. knowledge graph completion from listings, exploitation of listing context, as well as rule learning for knowledge graphs.

\subsection{Knowledge Graph Completion from Listings}
\label{subsec:knowledge-graph-completion-from-listings}
Knowledge graph completion using information in web tables has already been an active research area in the last several years. In 2016, Ritze et al. \cite{ritze2016profiling} profiled the potential of web tables in the WDC Web Table Corpus. Using the T2K Match framework, they match web tables to DBpedia and find that the best results for the extraction of new facts can be achieved using knowledge-based trust \cite{dong2015knowledge} (i.e., judging the quality of a set of extracted triples by their overlap with the knowledge base). Zhang et al. \cite{zhang2020novel} present an approach for detection of novel entities in tables. They first exploit lexical and semantic similarity for entity linking and column heading property matching. In a second step they use the output to detect novel entities in table columns. Oulabi and Bizer \cite{oulabi2019using} tackle the same problem for Wikipedia tables with a bootstrapping approach based on expert-defined rules. Macdonald and Barbosa \cite{macdonald2020neural} extract new facts from Wikipedia tables to extend the Freebase knowledge base. With an LSTM that uses contextual information of the table, they extract new facts for 28 relations.

Lists have only very sparsely been used for knowledge graph completion. Paulheim and Ponzetto \cite{paulheim2013extending} frame the general potential of list pages as a source of knowledge in Wikipedia. They propose to use a combination of statistical and NLP methods to extract knowledge and show that, by applying them to a single list page, they are able to extract a thousand new statements.

Compared to all previously mentioned approaches, we take an abstract view on listings by considering only their subject entities. This provides the advantage that rules can be learned from and applied to arbitrary listings. In addition to that, we do not only discover novel entities, but also discover relations between those entities and the page subject.

In our previous work \cite{heist2020entity}, we have already presented an approach for the identification of novel entities and the extraction of facts in Wikipedia list pages. List pages are pages in Wikipedia that start with \textit{List of} and contain listings (i.e., tables or lists) of entities for a given topic (e.g. \textit{List of Japanese speculative fiction writers}). The approach is divided into two phases: In a first phase, a dataset of tagged entities from list pages is extracted. With distant supervision from CaLiGraph, a knowledge graph with a detailed type hierarchy derived from Wikipedia categories and list pages, a part of the mentioned entities is heuristically labeled as subject entities and non-subject entities. In a second phase, the dataset is enriched with positional, lexical, and statistical features extracted from the list pages. On the basis of this data, an XGBoost classifier is able to identify more than two million subject entities with an average precision of 90\%. As not all the information about the subject entities is contained in the knowledge graphs DBpedia and CaLiGraph, they can be enhanced with the missing information.

In this work, we reuse the approach presented in \cite{heist2020entity} for identifying subject entities. Further, as it is the only approach that also works with arbitrary listings, we use it as a baseline in our experiments. As, in its current state, it only works for list pages in Wikipedia, we extend it to arbitrary pages with a simple frequency-based approach.

\subsection{Exploiting the Context of Listings}
As tables are the more actively researched type of listings, we focus here on the types of context used when working with tables. The most obvious source of context is found directly on the page where the table is located. This page context is, for example, used by InfoGather \cite{yakout2012infogather} to detect possible synonyms in table headers for means of table matching.

Zhang \cite{zhang2014towards} distinguishes between "in-table" features like the table header, and "out-table" features like captions, page title, and text of surrounding paragraphs. With both kinds of features, they perform entity disambiguation against Freebase.

The previously mentioned approach of Macdonald and Barbosa \cite{macdonald2020neural} focuses on tables in Wikipedia and hence uses specific context features like section titles, table headers and captions, and the text in the first paragraph of the table's section. Interestingly, they do not only discover relations between entities in the table, but also between a table entity and the page subject.

MENTOR \cite{cannaviccio2018leveraging} leverages patterns occurring in headers of Wikipedia tables to consistently discover DBpedia relations. Lehmberg et al. \cite{lehmberg2017stitching} tackle the problem of small web tables with table stitching, i.e., they combine several small tables with a similar context (e.g., same page or domain and a matching schema) into one large table, making it easier to extract facts from it.

Apart from page context, many approaches use the context of entities in tables to improve extraction results. Zhang et al. \cite{zhang2020generating} generate new sub-classes to a taxonomy for a set of entities. Therefore, they find the best-describing class using the context of the entities. In particular, they use the categories of the entities as well as the immediate context around the entities on the page. Another approach that uses entity categories as context is TableNet \cite{fetahu2019tablenet}. They leverage the context to find schematically similar or related tables for a given table in Wikipedia.

In our experiments with Wikipedia, we use section headers as page context and types in the knowledge graph as entity context. However, the definition of context in our approach is kept very generic on purpose. By doing that, we are able to incorporate additional context sources like section text or entity categories to improve extraction results. This, however, also comes with an increase in rule complexity and, consequently, run time.

\subsection{Rule-based Knowledge Graph Completion}
Rule-based knowledge graph completion approaches typically generate rules either on instance-level (rules that add new facts for individual instances) or on schema-level (rules that add additional schematic constraints).

AMIE+ \cite{galarraga2015fast} and AnyBURL \cite{meilicke2019anytime} are instance-level rule learners inspired by inductive logic programming (ILP). The former uses top-down, the latter bottom-up rule learning to generate rules in the fashion of $born(X,A) \land capital(A,Y) \implies citizen(X,Y)$.

DL-Learner \cite{lehmann2009dl} is an ILP-based approach on schema-level which finds description logic patterns for a set of instances. A related approach uses statistical schema induction \cite{volker2011statistical} to derive additional schema constraints (e.g. range restrictions for predicates).

The above mentioned approaches are merely \emph{link prediction} approaches, i.e. they predict new relations between entities already contained in the knowledge graph. The same holds for the omnipresent knowledge graph embedding approaches \cite{wang2017knowledge}. Such approaches are very productive when enough training data is available and they provide exact results especially when both positive and negative examples are given. In the setting of this paper, we are working with (more or less) noisy external data.

With regard to instance- versus schema-level, our approach can be regarded as a hybrid approach that generates rules for sets of entities, which are in turn used to generate facts on an instance-level. In this respect, our approach is similar to C-DF \cite{xu2016learning} which uses Wikipedia categories as an external data source to derive the characteristics of categories. To that end, they derive lexical patterns from category names and contained entities.

In this paper, we apply rule learning to co-occurring entities in Wikipedia. While existing approaches have only considered explicit co-occurrence, i.e., categories or list pages, we go beyond the state of the art by considering \emph{arbitrary} listings in Wikipedia, as the one shown in Fig.~\ref{fig:running-example}.

\section{Information Extraction From Co-Occurrences}
\label{sec:ie-from-cooccurrences}
In this paper, we consider a data corpus $D$ from which co-occurring entities can be extracted (e.g., listings in Wikipedia or a collection of spreadsheets). Furthermore, we assume that a knowledge graph which contains a subset of those entities can be extended with information learned about the co-occurring entities.
\subsection{Task Formulation}
\label{subsec:task-formulation}
The Knowledge Graph $\mathcal{K}$ is a set of assertions about its entities in the form of triples $\{(s,p,o)|s \in \mathcal{E}, p \in \mathcal{P}, o \in \mathcal{E} \cup \mathcal{T} \cup \mathcal{L}\}$ defined over sets of entities $\mathcal{E}$, predicates $\mathcal{P}$, types $\mathcal{T}$, and literals $\mathcal{L}$. We refer to statements about the types of an entity (i.e., $p = \textrm{rdf:type}, o \in \mathcal{T}$) as type assertions ($TA \subset K$), and to statements about relations between two entities (i.e., $o \in \mathcal{E}$) as relation assertions ($RA \subset K$). With $\mathcal{K}^* \supseteq \mathcal{K}$, we refer to the idealized complete version of $\mathcal{K}$. With regard to the OWA this means that a fact is incorrect if it is not contained in $\mathcal{K}^*$.\footnote{$\mathcal{K}^*$ is merely a theoretical construct, since a complete knowledge graph of all entities in the world cannot exist.}

The data corpus $D$ contains a set of listings $\Phi$, where each listing $\phi \in \Phi$ contains a number of subject entities $SE_\phi$. Our task is to identify statements that hold for all subject entities $SE_\phi$ in a listing $\phi$. We distinguish taxonomic and relational information that is expressed in $\mathcal{K}$.

The taxonomic information is a set of types that is shared by all SE of a listing:
\begin{equation}\label{eq:types-of-listing}
    \mathcal{T}_\phi = \{t | t \in \mathcal{T}, \forall s \in SE_\phi : (s, \mathtt{rdf}\mathord{:}\mathtt{type}, t) \in \mathcal{K}^*\},
\end{equation}

and the relational information is a set of relations to other entities which is shared by all SE of a listing:\footnote{Here, the entities in $SE_\phi$ may occur both in the subject as well as in the object position. But for a more concise notation, we use only (p,o)-tuples and introduce the set of inverse predicates $\mathcal{P}^{-1}$ to express that SE may also occur in object position. This is, however, only a notation and the inverse predicates do not have to exist in the schema.}
\begin{eqnarray}\label{eq:rels-of-listing}
    \mathcal{R}_\phi = \{(p,o) | p \in \mathcal{P} \cup \mathcal{P}^{-1}, o \in \mathcal{E}, \forall s \in SE_\phi : (s,p,o) \in \mathcal{K}^*\}.
\end{eqnarray}

From these characteristics of listings, we can derive all the \emph{additional} type assertions
\begin{equation}\label{eq:ta-plus}
    TA^+ = \bigcup_{\phi \in \Phi} \{(s, \mathtt{rdf}\mathord{:}\mathtt{type}, t)|s \in SE_\phi, t \in \mathcal{T}_\phi\} \setminus TA
\end{equation}

and \emph{additional} relation assertions
\begin{equation}\label{eq:ra-plus}
    RA^+ = \bigcup_{\phi \in \Phi} \{(s,p,o)|s \in SE_\phi, (p,o) \in \mathcal{R}_\phi\} \setminus RA
\end{equation}

that are encoded in $\Phi$ and missing in $\mathcal{K}$. Furthermore, $TA^+$ and $RA^+$ can contain additional entities that are not yet contained in $\mathcal{K}$, as there is no restriction for subject entities of $\Phi$ to be part of $\mathcal{K}$.

For the sake of readability, we will only describe the case of $\mathcal{R}_\phi$ for the remainder of this section as $\mathcal{T}_\phi$ is -- notation-wise -- a special case of $\mathcal{R}_\phi$ with $p = \mathtt{rdf}\textrm{:}\mathtt{type}$ and $o \in \mathcal{T}$.

\subsection{Learning Descriptive Rules for Listings}
\label{subsec:learning-descriptive-rules-for-listings}
Due to the incompleteness of $\mathcal{K}$, it is not possible to derive the exact set of relations $\mathcal{R}_\phi$ for every listing in $\Phi$. Hence, our goal is to derive an approximate version $\mathcal{\hat{R}}_\phi$ by using $\phi$ and the knowledge about $SE_\phi$ in $\mathcal{K}$.

Similar to the rule learner AMIE+ \cite{galarraga2015fast}, we use the partial completeness assumption (PCA) to generate negative evidence. The PCA implies that if $(s,p,o) \in \mathcal{K}$ then $\forall o' : (s,p,o') \in \mathcal{K}^* \implies (s,p,o') \in \mathcal{K}$. In order words, if $\mathcal{K}$ makes some assertions with a predicate $p$ for a subject $s$, then we assume that $\mathcal{K}$ contains every $p$-related information about $s$.

Following from the PCA, we use the $count$ of entities with a specific predicate-object combination in a set of entities $E$
\begin{equation}\label{eq:relation-count}
count(E,p,o) = |\{s|s \in E, \exists o : (s,p,o) \in \mathcal{K}\}|
\end{equation}

and the $count$ of entities having predicate $p$ with an arbitrary object
\begin{equation}\label{eq:predicate-count}
count(E,p) = |\{s|s \in E, \exists o' : (s,p,o') \in \mathcal{K}\}|
\end{equation}

to compute a maximum-likelihood-based frequency of a specific predicate-object combination occurring in $E$:
\begin{equation}\label{eq:relation-frequency}
freq(E,p,o) = \frac{count(E,p,o)}{count(E,p)}.
\end{equation}

From Eq.~\ref{eq:relation-frequency} we first derive a naive approximation of a listing's relations by including all relations with a frequency above a defined threshold $\tau_{freq}$:
\begin{equation}\label{eq:freq-r-hat}
    \mathcal{\hat{R}}_\phi^{freq} = \{(p,o) | (p,o) \in \mathcal{R}, freq(SE_\phi,p,o) > \tau_{freq}\}.
\end{equation}

As argued in Sec.~\ref{subsec:motivation-and-problem}, we improve this naive frequency-based approximation by learning more general patterns that describe the characteristics of listings using their context.

\textbf{Hypothesis 1.} The context $\zeta_\phi$ of a listing $\phi$ in $D$ contains such information about $\mathcal{R}_\phi$ that it can be used to find subsets of $\Phi$ with similar $\mathcal{R}$.

\begin{table}
  \caption{Exemplary context ($\zeta$), type frequency ($T^F$), and relation frequency ($R^F$) vectors for a set of listings extracted from $D$. While $\zeta$ is extracted directly from $D$, $T^F$ and $R^F$ are retrieved via distant supervision from $\mathcal{K}$.}
  \label{tab:task-listings}
  \begin{tabular}{cccc}
    \toprule
    Listing & $\zeta$ & $T^F$ & $R^F$\\
    \midrule
    $\phi_1$ & (1 0 1 ... 1) & (0.2 0.9 0.0 ... 0.1) & (0.9 0.1 0.0 ... 0.1)\\
    $\phi_2$ & (0 1 1 ... 0) & (0.0 0.2 0.0 ... 0.9) & (0.0 0.0 0.0 ... 0.2)\\
    $\phi_3$ & (0 0 0 ... 0) & (0.7 0.7 0.0 ... 0.0) & (0.0 0.0 0.0 ... 0.4)\\
    \multicolumn{4}{c}{...}\\
    $\phi_{n-1}$ & (1 0 0 ... 1) & (0.8 0.9 0.0 ... 0.0) & (0.0 0.9 0.0 ... 0.0)\\
    $\phi_n$ & (1 0 0 ... 1) & (0.7 1.0 0.0 ... 0.3) & (0.0 0.0 0.8 ... 0.0)\\
  \bottomrule
\end{tabular}
\end{table}

Let Table~\ref{tab:task-listings} contain the information about all listings in $D$. A listing $\phi$ is defined by its context $\zeta_\phi$ (which can in theory contain any information about $\phi$, from the title of its section to an actual image of the listing), the type frequencies $(t_1, t_2,.., t_x) \in T^F_\phi$, and the relation frequencies $(r_1, r_2,.., r_y) \in R^F_\phi$. Listings $\phi_1$, $\phi_{n-1}$, and $\phi_n$ have overlapping context vectors. $t_2$ has a consistently high frequency over all three listings. It is thus a potential type characteristic for this kind of listing context. Furthermore, $r_1$ has a high frequency in $\phi_1$, $r_2$ in $\phi_{n-1}$, and $r_3$ in $\phi_n$ -- if the three relations share the same predicate, they may all express a similar relation to an entity in their context (e.g. to the subject of the page). 

In a concrete scenario, the context vector \textit{(1 0 0 ... 1)} might indicate that the listing is located on the page of a musician under the section \textit{Solo albums}. $t_2$ holds the frequency of the type \textit{Album} in this listing and $r_1$ to $r_3$ describe the frequencies of the relations (\textit{artist}, Gilby Clarke), (\textit{artist}, Axl Rose), and (\textit{artist}, Slash).

We formulate the task of discovering frequent co-occurrences of context elements and taxonomic and relational patterns as an association rule mining task over all listings in $D$. Association rules, as introduced by Agrawal et al. \cite{agrawal1993mining}, are simple implication patterns originally developed for large and sparse datasets like transaction databases of supermarket chains. To discover items that are frequently bought together, rules of the form $X \implies Y$ are produced, with $X$ and $Y$ being itemsets. In the knowledge graph context, they have been used, e.g., for enriching the schema of a knowledge graph \cite{paulheim2012unsupervised,volker2011statistical}.

For our scenario, we need a mapping from a context vector $\zeta \in Z$ to a predicate-object tuple. Hence, we define a rule $r$, its antecedent $r_a$, and its consequent $r_c$ as follows:
\begin{equation}\label{eq:rule-definition}
    r: r_a \in Z \implies r_c \in (\mathcal{P} \cup \mathcal{P}^{-1})\times (\mathcal{T} \cup \mathcal{E} \cup \mathcal{X}).
\end{equation}
As a rule should be able to imply relations to entities that vary with the context of a listing (e.g. to \textit{Gilby Clarke} as the page's subject in Fig.~\ref{fig:running-example}), we introduce $\mathcal{X}$ as the set of placeholders for context entities (instead of \textit{Gilby Clarke}, the object of the rule's consequent would be <\texttt{PageEntity}>).

We say a rule antecedent $r_a$ matches a listing context $\zeta_\phi$ (short: $r_a \simeq \zeta_\phi$) if the vector of $\zeta_\phi$ is 1 when the vector of $r_a$ is 1. In essence, $\zeta_\phi$ must comprise $r_a$. Accordingly, we need to find a set of rules $R$, so that for every listing $\phi$ the set of approximate listing relations
\begin{equation}\label{eq:rule-t-hat}
    \mathcal{\hat{R}}_\phi^{rule} = \bigcup_{r \in R} \{r_c|r_a \simeq \zeta_\phi\}
\end{equation}
resembles the true relations $\mathcal{R}_\phi$ as closely as possible.

Considering all the listings in Fig.~\ref{fig:running-example}, their $\mathcal{\hat{R}}_\phi^{rule}$ should, among others, contain the rules\footnote{Note that Eqs.~\ref{eq:axiom-example-1} and \ref{eq:axiom-example-2} are the axiom equivalents of Eqs.~\ref{eq:rule-example-1} and \ref{eq:rule-example-2}. For better readability, we use the description logics notation of Eqs.~\ref{eq:axiom-example-1} and \ref{eq:axiom-example-2} from here on.}$^,$\footnote{Instead of a binary vector, we use a more expressive notation for the listing context in our examples. The notations are trivially convertible by one-hot-encoding.}
\begin{equation}\label{eq:rule-example-1}
topSection(\textrm{"Discography"}) \implies (type,\mathtt{MusicalWork})
\end{equation}
and
\begin{equation}\label{eq:rule-example-2}
topSection(\textrm{"Discography"}) \implies (artist,\textrm{<}PageEntity\textrm{>}).
\end{equation}

It is important to note that these rules can be derived from listings with differing context vectors. All listings only have to have in common that their top section has the title \textit{Discography} and that the contained entities are of the type \texttt{MusicalWork} with the page entity as artist. Still, the individual listings may, for example, occur in sections with different titles.

\subsection{Quality Metrics for Rules}
\label{subsec:quality-metrics-for-rules}
In original association rule mining, two metrics are typically considered to judge the quality of a rule $X \implies Y$: the support of the rule antecedent (how often does $X$ occur in the dataset), and the confidence of the rule (how often does $X \cup Y$ occur in relation to $X$).

Transferring the support metric to our task, we count the absolute frequency of a particular context occurring in $\Phi$. Let $\Phi_{r_a} = \{\phi|\phi \in \Phi, r_a \simeq \zeta_\phi\}$, then we define the support of the rule antecedent $r_a$ as
\begin{equation}\label{eq:pattern-support}
    supp(r_a) = |\Phi_{r_a}|.
\end{equation}

Due to the incompleteness of $\mathcal{K}$, the values of $Y$ are in our case no definitive items but maximum-likelihood estimates of types and relations. With respect to these estimates, a good rule has to fulfill two criteria: it has to be correct (i.e. frequent with respect to all SE of the covered listings) and it has to be consistent (i.e. consistently correct over all the covered listings).

We define the correctness, or confidence, of a rule as the frequency of the rule consequent over all SE of a rule's covered listings:
\begin{equation}\label{eq:pattern-confidence}
    conf(r) = \frac{\sum_{\phi \in \Phi_{r_a}} count(SE_\phi,p_{r_c},o_{r_c})}{\sum_{\phi \in \Phi_{r_a}} count(SE_\phi,p_{r_c})},
\end{equation}

and we define the consistency of a rule using the mean absolute deviation of an individual listing's confidence to the overall confidence of the rule:
\begin{equation}\label{eq:pattern-consistency}
    cons(r) = 1 - \frac{\sum_{\phi \in \Phi_{r_a}} |freq(SE_\phi,p_{r_c},o_{r_c}) - conf(r)|}{supp(r_a)}.
\end{equation}

While a high confidence ensures that the overall assertions generated by the rule are correct, a high consistency ensures that few listings with many SE do not outvote the remaining covered listings.

To select an appropriate set of rules $R$ from all the candidate rules $R^*$ in the search space, we have to pick reasonable thresholds for the minimum support ($\tau_{supp}$), the minimum confidence ($\tau_{conf}$), and the minimum consistency ($\tau_{cons}$). By applying these thresholds, we find our final set of descriptive rules $R$:
\begin{equation}\label{eq:rule-selection}
    \{r | r \in R^*, supp(r_a) > \tau_{supp} \land conf(r) > \tau_{conf} \land cons(r) > \tau_{cons}\}.
\end{equation}
Typically, the choice of these thresholds is strongly influenced by the nature of the dataset $D$ and the extraction goal (correctness versus coverage).

\section{Exploiting Co-Occurrences in Wikipedia}
\label{sec:exploiting-co-occurrences-in-Wikipedia}
Wikipedia is a rich source of listings, both in dedicated list pages as well as in sections of article pages. Hence, we use it as a data corpus for our experiments. In Sec.~\ref{sec:discussion-and-outlook}, we discuss other appropriate corpora for our approach.

Due to its structured and encyclopedic nature, Wikipedia is a perfect application scenario for our approach. We can exploit the structure by building very expressive context vectors. Obviously, this positively influences the quality of extraction results. Still, the definition of the context vector is kept abstract on purpose to make the approach applicable to other kinds of web resource as well. However, an empirical evaluation of the practicability or performance of the approach for resources outside of the encyclopedic domain is out of scope of this paper.

\subsection{Approach Overview}
\begin{figure*}[ht]
  \centering
  \includegraphics[width=.98\textwidth]{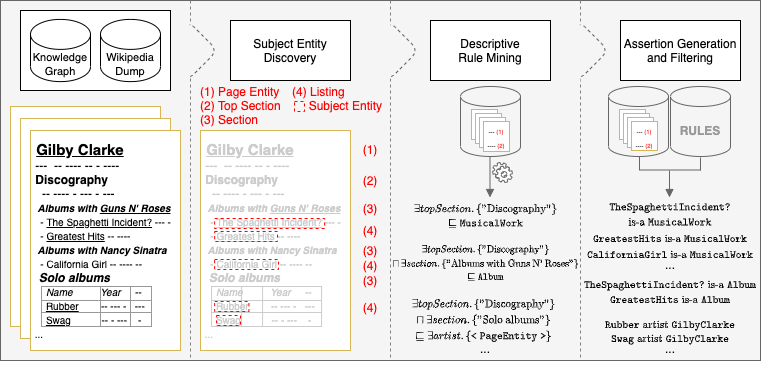}
  \caption{An overview of the approach with exemplary outputs of the individual phases.}
  \label{fig:approach-overview}
  \Description{An overview of the approach to extract type and relation assertions from Wikipedia listings. The approach starts with a knowledge graph and a Wikipedia dump and then runs the three phases Subject Entity Discovery, Descriptive Rule Mining, and Assertion Generation and Filtering.}
\end{figure*}

Fig.~\ref{fig:approach-overview} gives an overview of our extraction approach. The input of the approach is a dump of Wikipedia as well as an associated knowledge graph. In the \textit{Subject Entity Discovery} phase, listings and their context are extracted from the Wikipedia dump and subject entities are identified (Sec.~\ref{subsec:subject-entity-discovery}). Subsequently, the existing information in the knowledge graph is used to mine descriptive rules from the extracted listings (Sec.~\ref{subsec:descriptive-rule-mining}). Finally, the rules are applied to all the listings in Wikipedia in order to extract new type and relation assertions (Sec.~\ref{subsec:assertion-generation-and-filtering}).

\subsection{Wikipedia as a Data Corpus}
We pick Wikipedia as a data corpus for our experiments as it brings several advantages:

\paragraph{Structure} Wikipedia is written in an entity-centric style with a focus on facts. Listings are often used to provide an overview of a set of entities that are related to the main entity. Due to the encyclopedic style and the peer-reviewing process, it has a consistent structure. Especially section titles are used consistently for specific topics. Wikipedia has its own markup language (Wiki markup), which allows a more consistent access to interesting page structures like listings and tables than plain HTML.

\paragraph{Entity Links} If a Wikipedia article is mentioned in another article, it is typically linked in the Wiki markup (a so called \textit{blue link}). Furthermore, it is possible to link to an article that does not (yet) exist (a so called \textit{red link}). As Wikipedia articles can be trivially mapped to entities in Wikipedia-based knowledge graphs like DBpedia, since they create one entity per article, we can identify many named entities in listings and their context without the help of an entity linker.

For our experiments, we use a Wikipedia dump of October 2016 which is, at the time of the experiments, the most recent dump that is compatible with both DBpedia and CaLiGraph. In this version, Wikipedia contains 6.9M articles, 2.4M of which contain listings with at least two rows.\footnote{Wiki markup is parsed with WikiTextParser: \url{https://github.com/5j9/wikitextparser}.} In total, there are 5.1M listings with a row count median of 8, mean of 21.9, and standard deviation of 76.8. Of these listings, 1.1M are tables, and 4.0M are lists.

\subsection{Subject Entity Discovery}
\label{subsec:subject-entity-discovery}
\subsubsection{Entity Tagging}
Apart from the already tagged entities via blue and red links, we have to make sure that any other named entity in listings and their context is identified as well. This is done in two steps:

In a first step, we expand all the blue and red links in an article. If a piece of text is linked to another article, we make sure that every occurrence of that piece of text in the article is linked to the other article. This is necessary as by convention other articles are only linked at their first occurrence in the text.\footnote{\url{https://en.wikipedia.org/wiki/Wikipedia:Manual_of_Style/Linking\#Duplicate_and_repeat_links}}

In a second step, we use a named entity tagger to identify additional named entities in listings. To that end, we use a state-of-the-art entity tagger from spaCy.\footnote{\url{https://spacy.io}} This tagger is trained on the OntoNotes5\footnote{\url{https://catalog.ldc.upenn.edu/LDC2013T19}} corpus, and thus not specifically trained to identify named entities in short text snippets like they occur in listings. Therefore, we specialize the tagger by providing it Wikipedia listings as additional training data with blue links as positive examples. In detail, the tagger is specialized as follows:
\begin{itemize}
    \item We retrieve all listings in Wikipedia list pages as training data.
    \item We apply the plain spaCy entity tagger to the listings to get named entity tags for all mentioned entities.
    \item To make these tags more consistent, we use information from DBpedia about the tagged entities: We look at the distribution of named entity tags over entities with respect to their DBpedia types and take the majority vote. For example, if 80\% of entities with the DBpedia type \texttt{Person} are annotated with the tag \textit{PERSON}, we use \textit{PERSON} as label for all these entities.
    \item Using these consistent named entity tags for blue-link entities, we specialize the spaCy tagger.
\end{itemize}

\subsubsection{Subject Entity Classification}
We apply the approach from \cite{heist2020entity} for the identification of subject entities in listings. In short, we use lexical, positional, and statistical features to classify entities as subject or non-subject entities (refer to Sec.~\ref{subsec:knowledge-graph-completion-from-listings} for more details). Despite being developed only for listings in list pages, the classifier is applicable to any kind of listing in Wikipedia. A disadvantage of this broader application is that the classifier is not trained in such a way that it ignores listings used for organisational or design purposes (e.g. summaries or timelines). These have to be filtered out in the subsequent stages.

\subsubsection{Results}
After expanding all the blue and red links on the pages, the dataset contains 5.1M listings with 60.1M entity mentions. 51.6M additional entity mentions are identified by the named entity tagger.

Of all the entity mentions, we classify 25.8M as subject entities. Those occur in 2.5M listings of 1.3M pages. This results in a mean of 10.5 and median of 4 subject entities per listing with a standard deviation of 49.8.

\subsection{Descriptive Rule Mining}
\label{subsec:descriptive-rule-mining}
\subsubsection{Describing Listings}
The search space for rule candidates is defined by the listing context. Thus, we choose the context in such a way that it is expressive enough to be an appropriate indicator for $\mathcal{T}_\phi$ and $\mathcal{R}_\phi$, and concise enough to explore the complete search space without any additional heuristics.

We exploit the fact that Wikipedia pages of a certain type (e.g., musicians) mostly follow naming conventions for the sections of their articles (e.g., albums and songs are listed under the top section \textit{Discography}). Further, we exploit that the objects of the SE's relations are usually either the entity of the page, or an entity mentioned in a section title. We call these typical places for objects the relation \textit{targets}. In Fig.~\ref{fig:running-example}, \textit{Gilby Clarke} is an example of a \textit{PageEntity} target, and \textit{Guns N' Roses} as well as \textit{Nancy Sinatra} are examples for \textit{SectionEntity} targets. As a result, we use the type of the page entity, the top section title, and the section title as listing context.

Additionally, we use the type of entities that are mentioned in section titles. This enables the learning of more abstract rules, e.g., to distinguish between albums listed in a section describing a band:
\begin{gather*}
\exists pageEntityType.\{\mathtt{Person}\} \sqcap \exists topSection.\{\textrm{"Discography"}\}\\
\sqcap \exists sectionEntityType.\{\mathtt{Band}\} \sqsubseteq \mathtt{Album},
\end{gather*}

and songs listed in a section describing an album:
\begin{gather*}
\exists pageEntityType.\{\mathtt{Person}\} \sqcap \exists topSection.\{\textrm{"Discography"}\}\\
\sqcap \exists sectionEntityType.\{\mathtt{Album}\} \sqsubseteq \mathtt{Song}.
\end{gather*}

\subsubsection{Threshold Selection}\label{subsubsec:threshold-selection}
We want to pick the thresholds in such a way that we tolerate some errors and missing information in $\mathcal{K}$, but do not allow many over-generalized rules that create incorrect assertions. Our idea for a sensible threshold selection is based on two assumptions:

\textbf{Assumption 1.} Being based on a maximum-likelihood estimation, rule confidence and consistency roughly order rules by the degree of prior knowledge we have about them.

\textbf{Assumption 2.} Assertions generated by over-generalized rules contain substantially more random noise than assertions generated by good rules.

Assumption~1 implies that the number of over-generalized rules increases with the decrease of confidence and consistency. As a consequence, assumption~2 implies that the amount of random noise increases with decrease of confidence and consistency.

To measure the increase of noise in generated assertions, we implicitly rely on existing knowledge in $\mathcal{K}$ by using the named entity tags of subject entities as a proxy. This works as follows: For a subject entity $e$ that is contained in $\mathcal{K}$, we have its type information $\mathcal{T}_e$ from $\mathcal{K}$ and we have its named entity tag $\psi_e$ from our named entity tagger. Going over all SE of listings in $\Phi$, we compute the probability of an entity with type $t$ having the tag $\psi$ by counting how often they co-occur:

\begin{equation}\label{eq:tag-probabilty}
    tagprob(t,\psi) = \frac{|\{e|\exists \phi \in \Phi : e \in SE_\phi \land t \in \mathcal{T}_e \land \psi = \psi_e\}|}{|\{e|\exists \phi \in \Phi : e \in SE_\phi \land t \in \mathcal{T}_e\}|}.
\end{equation}

For example, for the DBpedia type \texttt{Album}, we find the tag probabilities\\
\textit{WORK\_OF\_ART}: 0.49, \textit{ORG}: 0.14, \textit{PRODUCT}: 0.13, \textit{PERSON}: 0.07, showing that album titles are rather difficult to recognize. For the type \texttt{Person} and the tag \textit{PERSON}, on the other hand, we find a probability of 0.86.

We can then compute the tag-based probability for a set of assertions $A$ by averaging over the tag probability that is produced by the individual assertions. To compute this metric, we compare the tag of the assertion's subject entity with some kind of type information about it. This type information is either the asserted type (in case of a type assertion), or the domain of the predicate\footnote{We use the domain of the predicate $p$ as defined in $\mathcal{K}$. In case of $p \in \mathcal{P}^{-1}$, we use the range of the original predicate.} (in case of a relation assertion):
\begin{equation}\label{eq:tag-fit}
    tagfit(A) = 
    \begin{cases}
        \frac{\sum_{(s,p,o) \in A} tagprob(o,\psi_s)}{|A|} & \text{if } p = \mathtt{rdf}\mathord{:}\mathtt{type}\\
        \frac{\sum_{(s,p,o) \in A} tagprob(domain_p,\psi_s)}{|A|} & \text{otherwise.}\\
    \end{cases}
\end{equation}

While we do not expect the named entity tags to be perfect, our approach is based on the idea that the tags are consistent to a large extent. By comparing the $tagfit$ of assertions produced by rules with varying levels of confidence and consistency, we expect to see a clear decline as soon as too many noisy assertions are added.

\subsubsection{Results}
\begin{figure*}[ht]
\centering
\begin{subfigure}{.25\textwidth}
  \centering
  \includegraphics[width=.95\linewidth]{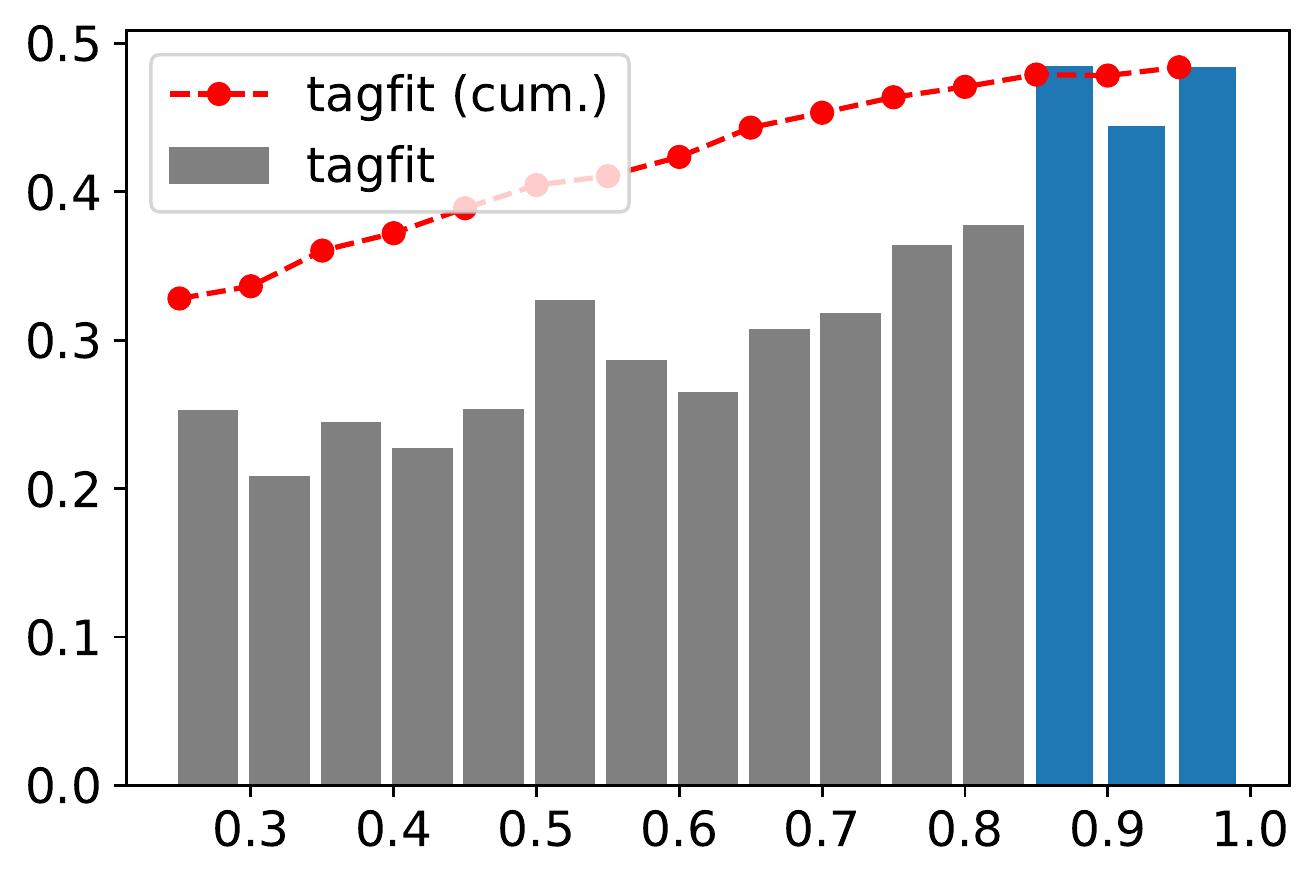}
  \caption{Type confidence}
  \label{fig:type-confidence}
\end{subfigure}%
\begin{subfigure}{.25\textwidth}
  \centering
  \includegraphics[width=.95\linewidth]{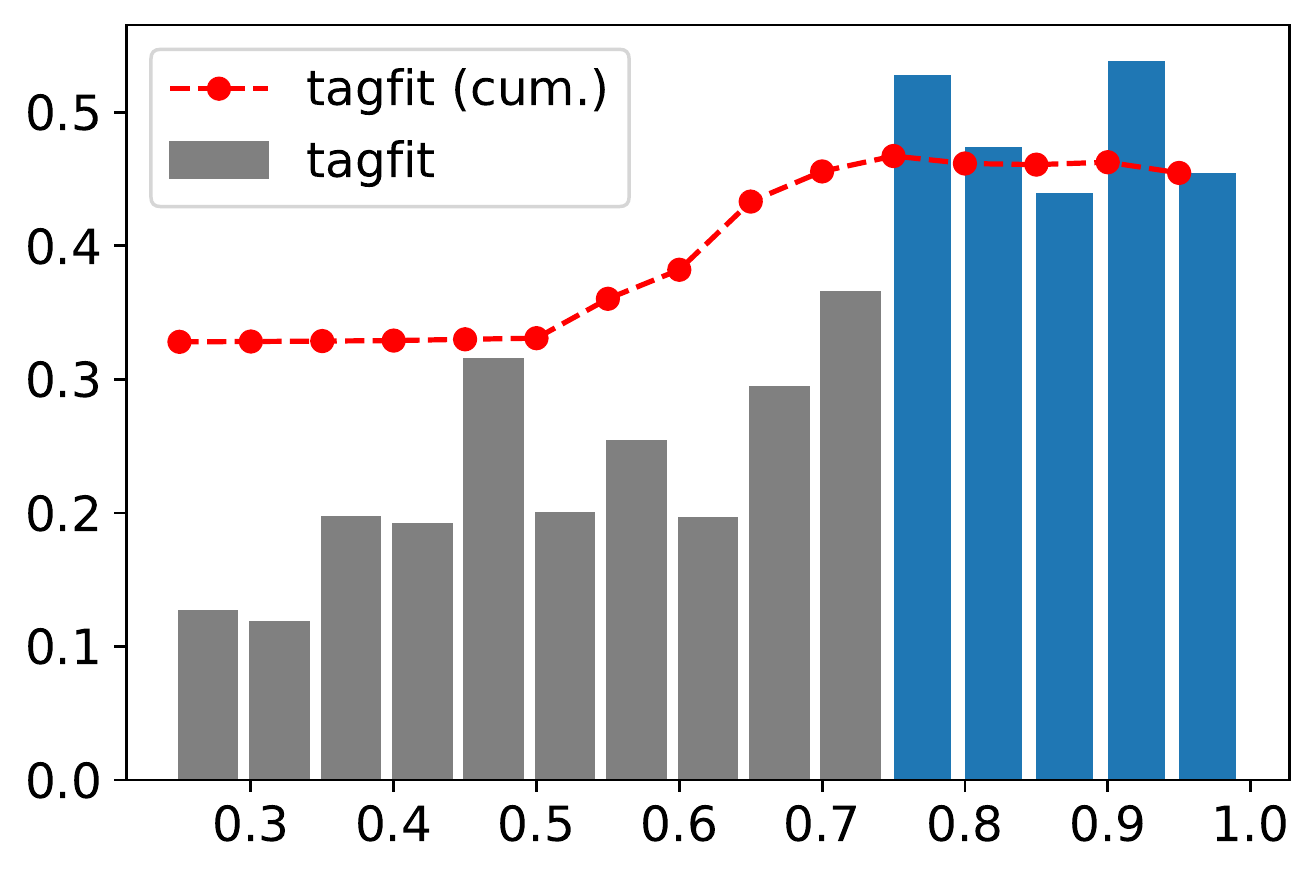}
  \caption{Type consistency}
  \label{fig:type-consistency}
\end{subfigure}%
\begin{subfigure}{.25\textwidth}
  \centering
  \includegraphics[width=.95\linewidth]{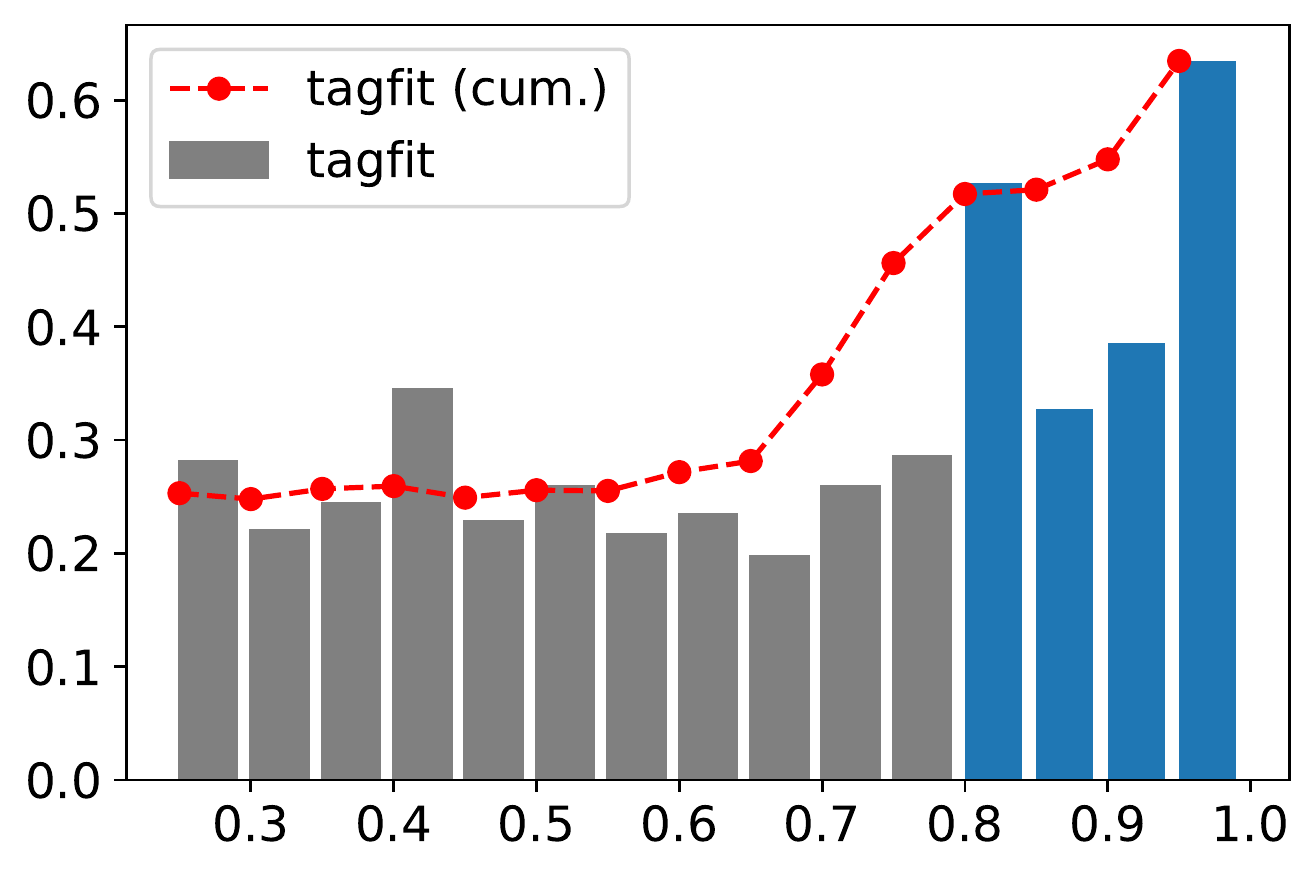}
  \caption{Relation confidence}
  \label{fig:relation-confidence}
\end{subfigure}%
\begin{subfigure}{.25\textwidth}
  \centering
  \includegraphics[width=.95\linewidth]{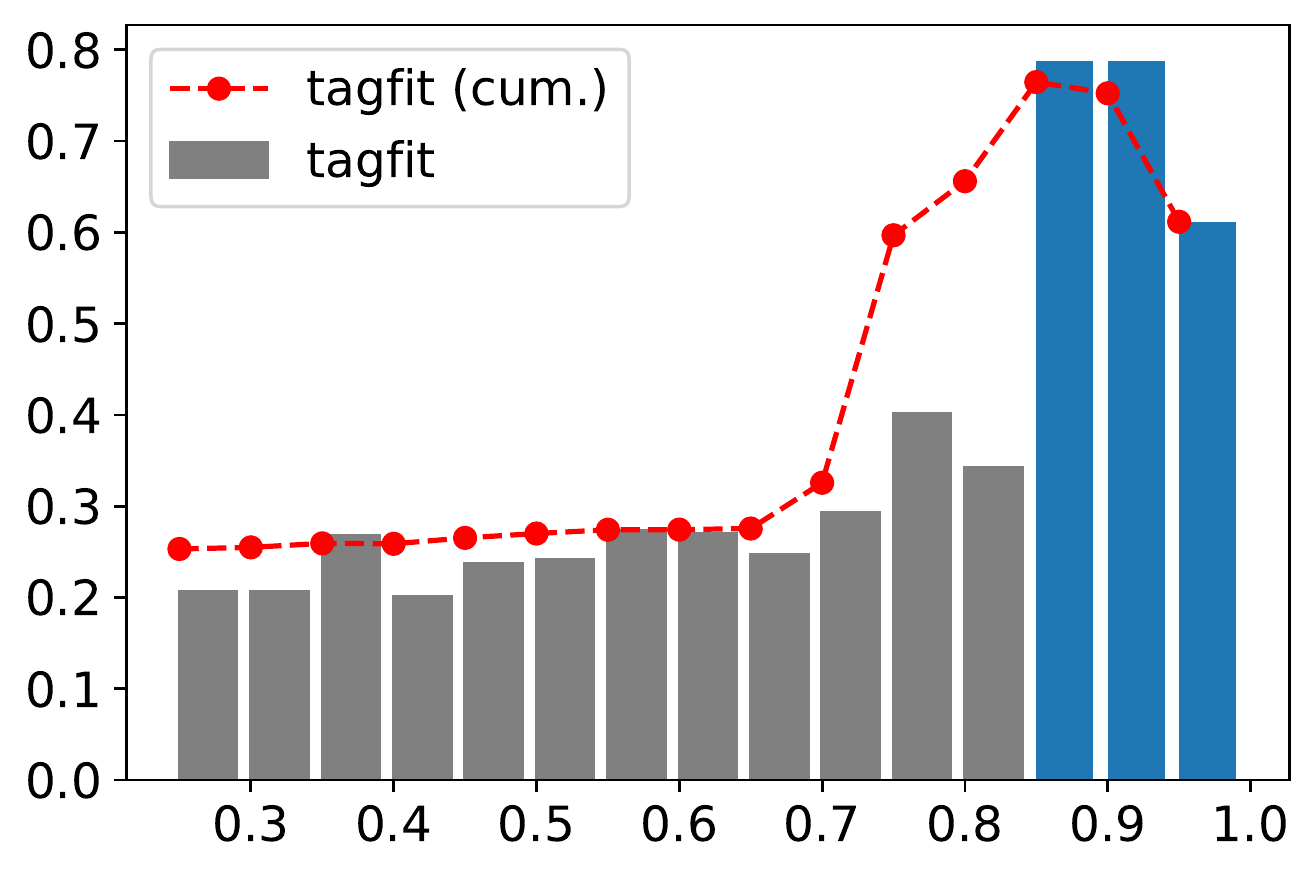}
  \caption{Relation consistency}
  \label{fig:relation-consistency}
\end{subfigure}
\caption{$tagfit$ of assertions generated from rules in a specified confidence or consistency interval. Bars show scores for a given interval (e.g. \textit{(0.75,0.80]}), lines show cumulative scores (e.g. \textit{(0.75,1.00]}). Blue bars indicate the selected threshold.}
\label{fig:thresholds}
\Description{Exact and cumulative scores for assertions of rules in a specified confidence or consistency interval. A first strong decrease of the tagfit can be seen at 0.85 and 0.75 for type confidence and consistency, and at 0.8 and 0.85 for relation confidence and consistency.}
\end{figure*}

Fig.~\ref{fig:thresholds} shows the $tagfit$ for type and relation assertions generated with varying levels of rule confidence and consistency. Our selection of thresholds is indicated by blue bars, i.e. we set the thresholds to the points where the $tagfit$ has its steepest drop. The thresholds are picked conservatively to select only high-quality rules by selecting points before an accelerated decrease of cumulative $tagfit$. But more coverage-oriented selections are also possible. In Fig.~\ref{fig:relation-consistency}, for example, a threshold of \textit{0.75} is also a valid option.

An analysis of rules with different levels of confidence and consistency has shown that a minimum support for types is not necessary. For relations, a support threshold of 2 is helpful to discard over-generalized rules. Further, we found that it is acceptable to pick the thresholds independently from each other, as the turning points for a given metric don't vary significantly when varying the remaining metrics.

Applying these thresholds, we find an overall number of 5,294,921 type rules with 369,139 distinct contexts and 244,642 distinct types. Further, we find 3,028 relation rules with 2,602 distinct contexts and 516 distinct relations. 949 of the relation rules have the page entity as target, and 2,079 have a section entity as target.

Among those rules are straightforward ones like
\begin{gather*}
\exists pageEntityType.\{\mathtt{Person}\} \sqcap \exists topSection.\{\textrm{"Acting filmography"}\}\\
\sqsubseteq \exists actor.\{\textrm{<}PageEntity\textrm{>}\},
\end{gather*}

and more specific ones like
\begin{gather*}
\exists pageEntityType.\{\mathtt{Location}\} \sqcap \exists topSection.\{\textrm{"Media"}\}\\
\sqcap \exists section.\{\textrm{"Newspapers"}\} \sqsubseteq \mathtt{Periodical\_literature}.
\end{gather*}

\subsection{Assertion Generation and Filtering}
\label{subsec:assertion-generation-and-filtering}
\subsubsection{Assertion Generation}
We apply the rules selected in the previous section to the complete dataset of listings to generate type and relation assertions. Subsequently, we remove any duplicate assertions and assertions that already exist in $\mathcal{K}$.

\subsubsection{Tag-based Filtering}
To get rid of errors introduced during the extraction process (e.g. due to incorrectly extracted subject entities or incorrect rules), we employ a final filtering step for the generated assertions: every assertion producing a $tagprob \leq \frac{1}{3}$ is discarded. The rationale behind the threshold is as follows: Types have typically one and sometimes two corresponding named entity tags (e.g. the tag \textit{PERSON} for the DBpedia type \texttt{Person}, or the tags \textit{ORG} and \textit{FAC} for the type \texttt{School}). As tag probabilities are relative frequencies, we make sure that, with a threshold of $\frac{1}{3}$, at most two tags are accepted for any given type.

For the tag probabilities of type \texttt{Album} from Sec.~\ref{subsubsec:threshold-selection}, the only valid tag is \textit{WORK\_OF\_ART}. As a consequence, any assertions of the form $(s, rdf\textrm{:}type, \mathtt{Album})$ with $s$ having a tag other than \textit{WORK\_OF\_ART} are discarded.

\subsubsection{Results}
Tab.~\ref{tab:results-assertions} shows the number of generated type and relation assertions before and after the tag-based filtering. The number of inferred types are listed separately for DBpedia and CaLiGraph. For relations, we show two kinds: The entry \textit{Relations} lists the number of extracted assertions from rules. As DBpedia and CaLiGraph share the same set of predicates, these assertions are applicable to both graphs. Furthermore, as \textit{Relations (via CaLiGraph)}, we list the number of relations that can be inferred from the extracted CaLiGraph types via restrictions in the CaLiGraph ontology. CaLiGraph contains more than 300k of such restrictions that imply a relation based on a certain type. For example, the ontology contains the value restriction $$\mathtt{Pop\_rock\_song} \sqsubseteq \exists genre.\{\textrm{Pop music}\}.$$ As we extract the type \texttt{Pop\_rock\_song} for the Beach Boys song \textit{At My Window}, we infer the fact $(\textrm{At My Window}, genre, \textrm{Pop music})$.

For CaLiGraph, we find assertions for 3.5M distinct subject entities with 3M of them not contained in the graph. For DBpedia, we find assertions for 3.1M distinct subject entities with 2.9M of them not contained. The unknown subject entities are, however, not disambiguated yet. Having only small text snippets in listings as information about these entities, a disambiguation with general-purpose disambiguation approaches \cite{zhu2018exploiting} is not practical. We thus leave this as an own research topic for future work. For an estimation of the actual number of novel entities, we rely on previous work \cite{heist2020entity}, where we analyzed the overlap for red links in list pages. In that paper, we estimate an overlap factor of 1.07 which would -- when applied to our scenario -- reduce the number of actual novel entities to roughly 2.8M for CaLiGraph and 2.7M for DBpedia. In relation to the current size of those graphs, this would be an increase of up to 38\% and 54\%, respectively \cite{heist2020knowledge}.

\begin{table}
  \caption{Number of generated assertions after removing existing assertions (Raw), and after applying tag-based filtering (Filtered).}
  \label{tab:results-assertions}
  \begin{tabular}{lccc}
    \toprule
    \textbf{Assertion Type} & \textbf{Raw} & \textbf{Filtered}\\
    \midrule
    Types (DBpedia) & 11,459,047 & \hphantom{0}7,721,039\\
    Types (CaLiGraph) & 47,249,624 & 29,128,677\\
    \midrule
    Relations & \hphantom{00,}732,820 & \hphantom{00,}542,018\\
    Relations (via CaLiGraph) & \hphantom{0}1,381,075 & \hphantom{00,}796,910\\
  \bottomrule
\end{tabular}
\end{table}

\section{Evaluation}

\label{sec:evaluation}
\begin{table}
  \caption{Correctness of manually evaluated assertions.}
  \label{tab:assertion-evaluation-results}
  \begin{tabular}{lccc}
    \toprule
    \textbf{Assertion Type} & \textbf{\#Dataset} & \textbf{\#Samples} & \textbf{Correct [\%]}\\
    \midrule
    \textit{Types (DBpedia)}&&\\
    frequency-based & \hphantom{0}6,680,565 & \hphantom{0,}414 & 91.55 $\pm$ 2.68\\
    rule-based & \hphantom{0}7,721,039 & \hphantom{0,}507 & 93.69 $\pm$ 2.12\\
    \midrule
    \textit{Types (CaLiGraph)}&&\\
    frequency-based & 26,676,191 & 2,000 & 89.40 $\pm$ 1.23\\
    rule-based & 29,128,677 & 2,000 & 91.95 $\pm$ 1.19\\
    \midrule
    \textit{Relations}&&\\
    frequency-based & \hphantom{00,}392,673 & 1,000 & 93.80 $\pm$ 1.49\\
    rule-based & \hphantom{00,}542,018 & 1,000 & 95.90 $\pm$ 1.23\\
  \bottomrule
\end{tabular}
\end{table}

In our performance evaluation, we judge the quality of generated assertions from our rule-based approach. As a baseline, we additionally evaluate assertions generated by the frequency-based approach (see Eq.~\ref{eq:freq-r-hat}). For the latter, we use a threshold comparable to our rule-based approach (i.e., we set $\tau_{freq}$ to $\tau_{conf}$ and disregard listings with less than three subject entities).

\subsection{Evaluation Procedure}
The evaluated assertions are created with a stratified random sampling strategy. The assertions are thus distributed proportionally over all page types (like \texttt{Person} or \texttt{Place}) and sampled randomly within these.

The labeling of the assertions is performed by the authors with the procedure as follows: For a given assertion, first the page of the listing is inspected, then -- if necessary and available -- the page of the subject entity. If a decision cannot be made based on this information, a search engine is used to evaluate the assertion. Samples of the rule-based and frequency-based approaches are evaluated together and in random order to ensure objectivity.

Tab.~\ref{tab:assertion-evaluation-results} shows the results of the performance evaluation. In total, we evaluated 2,000 examples per approach for types and 1,000 examples per approach for relations. The taxonomy of CaLiGraph comprises the one of DBpedia. Thus, we evaluated the full sample for CaLiGraph types and report the numbers for both graphs, which is the reason why the sample size for DBpedia is lower. For relations, we only evaluate the ones that are generated directly from rules and not the ones inferred from CaLiGraph types, as the correctness of the inferred relations directly depends on the correctness of CaLiGraph types.

\subsection{Type and Relation Extraction}
The evaluation results in Tab.~\ref{tab:assertion-evaluation-results} show that the information extracted from listings in Wikipedia is of an overall high quality. The rule-based approach yields a larger number of assertions with a higher correctness for both types and relations.

For both approaches, the correctness of the extracted assertions is substantially higher for DBpedia. The reason for that lies in the differing granularity of knowledge graph taxonomies. DBpedia has 764 different types while CaLiGraph has 755,441 with most of them being more specific extensions of DBpedia types. For example, DBpedia might describe a person as \texttt{Athlete}, while CaLiGraph describes it as \texttt{Olympic\_field\_hockey\_player\_of\_South\_Korea}. The average depth of predicted types is 2.06 for the former and 3.32 for the latter.

While the asserted types are very diverse (the most predicted type is \texttt{Agent} with 7.5\%), asserted relations are dominated by the predicate \textit{genus} with 69.8\% followed by \textit{isPartOf} (4.4\%) and \textit{artist} (3.2\%). This divergence cannot be explained with a different coverage: In DBpedia, 72\% of entities with type \texttt{Species} have a \textit{genus}, and 69\% of entities with type \texttt{MusicalWork} have an \textit{artist}. But we identify two other influencing factors: Wikipedia has very specific guidelines for editing species, especially with regard to standardization and formatting rules.\footnote{\url{https://species.wikimedia.org/wiki/Help:General_Wikispecies}} In addition to that, the \textit{genus} relation is functional and hence trivially fulfilling the PCA. As our approach is strongly relying on this assumption and it potentially inhibits the mining of practical rules for non-functional predicates (like, for example, for \textit{artist}), we plan on investigating this relationship further.

The inferred relations from CaLiGraph types are not evaluated explicitly. However, based on the correctness of restrictions in CaLiGraph that is reported to be 95.6\% \cite{heist2019uncovering} and from the correctness of type assertions, we estimate the correctness of the resulting relation assertions to be around 85.5\% for the frequency-based and around 87.9\% for the rule-based approach.

\subsection{Novel Entity Discovery}
\label{subsubsec:novel-entity-discovery}
For CaLiGraph, the frequency-based approach finds assertions for 2.5M distinct subject entities (2.1M of them novel). While the rule-based approach finds 9\% more assertions, its assertions are distributed over 40\% more entities (and over 43\% more novel entities). This demonstrates the capabilities of the rule-based approach to apply contextual patterns to environments where information about actual entities is sparse.

Further, we analyzed the portion of evaluated samples that applies to novel entities and found that the correctness of these statements is slightly better (between 0.1\% and 0.6\%) than the overall correctness. Including CaLiGraph types, we find an average of 9.03 assertions per novel entity, with a median of 7. This is, again, due to the very fine-grained type system of CaLiGraph. For example, for the rapper \textit{Dizzle Don}, which is a novel entity, we find 8 types (from \texttt{Agent} over \texttt{Musician} to \texttt{American\_rapper}) and 4 relations: (\textit{occupation}, Singing), (\textit{occupation}, Rapping), (\textit{birthPlace}, United States), and (\textit{genre}, Hip hop music).

\subsection{Error Analysis}
With Tab.~\ref{tab:assertion-error-types}, we provide an analysis of error type frequencies for the rule-based approach on the basis of the evaluated sample. (1) is caused by the entity linker, mostly due to incorrect entity borders. For example, the tagger identifies only a part of an album title. (2) is caused by errors of the subject entity identification approach, e.g. when the approach identifies the wrong column of a table as the one that holds subject entities. (3) can have multiple reasons, but most often the applied rule is over-generalized (e.g. implying \texttt{Football\_player} when the listing is actually about athletes in general) or applied to the wrong listing (i.e., the context described by the rule is not expressive enough). Finally, (4) happens, for example, when a table holds the specifications of a camera as this cannot be expressed with the given set of predicates in DBpedia or CaLiGraph.

Overall, most of the errors are produced by incorrectly applied rules. This is, however, unavoidable to a certain extent as knowledge graphs are not error-free and the data corpus is not perfect. A substantial portion of errors is also caused by incorrectly parsed or identified subject entities. Reducing these errors can also have a positive impact on the generated rules as correct information about entities is a requirement for correct rules.

\begin{table}
  \caption{Error types partitioned by cause. The occurrence values are given as their relative frequency (per 100) in the samples evaluated in Tab.~\ref{tab:assertion-evaluation-results}.}
  \label{tab:assertion-error-types}
  \begin{tabular}{lcc}
    \toprule
    Error type&Type&Relation\\
    \midrule
    \textbf{(1) Entity parsed incorrectly} & 2.6 & 0.2\\
    \textbf{(2) Wrong subject entity identified} & 1.4 & 1.6\\
    \textbf{(3) Rule applied incorrectly} & 3.7 & 2.3\\
    \textbf{(4) Semantics of listing too complex} & 0.3 & 0.0\\
  \bottomrule
\end{tabular}
\end{table}

\section{Discussion and Outlook}
\label{sec:discussion-and-outlook}
In this work, we demonstrate the potential of exploiting co-occurring similar entities for information extraction, and especially for the discovery of novel entities. We show that it is possible to mine expressive descriptive rules for listings in Wikipedia which can be used to extract information about millions of novel entities.

To improve our approach, we are investigating more sophisticated filtering approaches for the generated assertions to reduce the margin from raw to filtered assertions (see Tab.~\ref{tab:results-assertions}). Furthermore, we are experimenting with more expressive rules (e.g. by including additional context like substring patterns or section text) to improve our Wikipedia-based approach.

At the moment, we extract entities from single pages. While entity disambiguation on single pages is quite simple (on a single Wikipedia page, it is unlikely that the same surface form refers to different entities), the disambiguation of entities across pages is a much more challenging problem. Here, entity matching across pages is required, which should, ideally, combine signals from the source pages as well as constraints from the underlying ontology.

Furthermore, we work towards applying our approach to additional data corpora. Since the only language-dependent ingredient of our approach is the named entity tagging, and the entity tagger we use in our experiments has models for various languages,\footnote{\url{https://spacy.io/models}} our approach can also be extended to various language editions of Wikipedia.

Besides Wikipedia, we want to apply the approach to wikis in the Fandom\footnote{\url{https://www.fandom.com/}} universe containing more than 380k wikis on various domains (among them many interesting wikis for our approach, like for example WikiLists\footnote{\url{https://list.fandom.com/wiki/Main_Page}}). For background knowledge, we plan to rely on existing knowledge graphs in this domain like DBkWik \cite{hertling2020dbkwik} or TiFi \cite{chu2019tifi}. In the longer term, we want to extend the applicability of the approach towards arbitrary web pages, using microdata and RDFa annotations \cite{meusel2014webdatacommons} as hooks for background knowledge.

\bibliographystyle{ACM-Reference-Format}
\bibliography{bibliography}


\begin{thebibliography}{39}


\ifx \showCODEN    \undefined \def \showCODEN     #1{\unskip}     \fi
\ifx \showDOI      \undefined \def \showDOI       #1{#1}\fi
\ifx \showISBNx    \undefined \def \showISBNx     #1{\unskip}     \fi
\ifx \showISBNxiii \undefined \def \showISBNxiii  #1{\unskip}     \fi
\ifx \showISSN     \undefined \def \showISSN      #1{\unskip}     \fi
\ifx \showLCCN     \undefined \def \showLCCN      #1{\unskip}     \fi
\ifx \shownote     \undefined \def \shownote      #1{#1}          \fi
\ifx \showarticletitle \undefined \def \showarticletitle #1{#1}   \fi
\ifx \showURL      \undefined \def \showURL       {\relax}        \fi
\providecommand\bibfield[2]{#2}
\providecommand\bibinfo[2]{#2}
\providecommand\natexlab[1]{#1}
\providecommand\showeprint[2][]{arXiv:#2}

\bibitem[\protect\citeauthoryear{Agrawal, Imieli{\'n}ski, and Swami}{Agrawal
  et~al\mbox{.}}{1993}]%
        {agrawal1993mining}
\bibfield{author}{\bibinfo{person}{Rakesh Agrawal}, \bibinfo{person}{Tomasz
  Imieli{\'n}ski}, {and} \bibinfo{person}{Arun Swami}.}
  \bibinfo{year}{1993}\natexlab{}.
\newblock \showarticletitle{Mining association rules between sets of items in
  large databases}. In \bibinfo{booktitle}{\emph{1993 ACM SIGMOD international
  conference on Management of data}}. \bibinfo{pages}{207--216}.
\newblock


\bibitem[\protect\citeauthoryear{Cannaviccio, Ariemma, Barbosa, and
  Merialdo}{Cannaviccio et~al\mbox{.}}{2018}]%
        {cannaviccio2018leveraging}
\bibfield{author}{\bibinfo{person}{Matteo Cannaviccio},
  \bibinfo{person}{Lorenzo Ariemma}, \bibinfo{person}{Denilson Barbosa}, {and}
  \bibinfo{person}{Paolo Merialdo}.} \bibinfo{year}{2018}\natexlab{}.
\newblock \showarticletitle{Leveraging {W}ikipedia table schemas for knowledge
  graph augmentation}. In \bibinfo{booktitle}{\emph{21st International Workshop
  on the Web and Databases}}. \bibinfo{pages}{1--6}.
\newblock


\bibitem[\protect\citeauthoryear{Chu, Razniewski, and Weikum}{Chu
  et~al\mbox{.}}{2019}]%
        {chu2019tifi}
\bibfield{author}{\bibinfo{person}{Cuong~Xuan Chu}, \bibinfo{person}{Simon
  Razniewski}, {and} \bibinfo{person}{Gerhard Weikum}.}
  \bibinfo{year}{2019}\natexlab{}.
\newblock \showarticletitle{TiFi: Taxonomy Induction for Fictional Domains}. In
  \bibinfo{booktitle}{\emph{The World Wide Web Conference}}.
  \bibinfo{pages}{2673--2679}.
\newblock


\bibitem[\protect\citeauthoryear{Del~Corro and Gemulla}{Del~Corro and
  Gemulla}{2013}]%
        {del2013clausie}
\bibfield{author}{\bibinfo{person}{Luciano Del~Corro} {and}
  \bibinfo{person}{Rainer Gemulla}.} \bibinfo{year}{2013}\natexlab{}.
\newblock \showarticletitle{Clausie: clause-based open information extraction}.
  In \bibinfo{booktitle}{\emph{The World Wide Web Conference}}.
  \bibinfo{pages}{355--366}.
\newblock


\bibitem[\protect\citeauthoryear{Dong, Gabrilovich, Murphy, Dang, Horn,
  Lugaresi, Sun, and Zhang}{Dong et~al\mbox{.}}{2015}]%
        {dong2015knowledge}
\bibfield{author}{\bibinfo{person}{Xin~Luna Dong}, \bibinfo{person}{Evgeniy
  Gabrilovich}, \bibinfo{person}{Kevin Murphy}, \bibinfo{person}{Van Dang},
  \bibinfo{person}{Wilko Horn}, \bibinfo{person}{Camillo Lugaresi},
  \bibinfo{person}{Shaohua Sun}, {and} \bibinfo{person}{Wei Zhang}.}
  \bibinfo{year}{2015}\natexlab{}.
\newblock \showarticletitle{Knowledge-{B}ased {T}rust: Estimating the
  Trustworthiness of Web Sources}.
\newblock \bibinfo{journal}{\emph{VLDB Endowment}} \bibinfo{volume}{8},
  \bibinfo{number}{9} (\bibinfo{year}{2015}), \bibinfo{pages}{938--949}.
\newblock


\bibitem[\protect\citeauthoryear{F{\"a}rber, Rettinger, and
  El~Asmar}{F{\"a}rber et~al\mbox{.}}{2016}]%
        {farber2016emerging}
\bibfield{author}{\bibinfo{person}{Michael F{\"a}rber}, \bibinfo{person}{Achim
  Rettinger}, {and} \bibinfo{person}{Boulos El~Asmar}.}
  \bibinfo{year}{2016}\natexlab{}.
\newblock \showarticletitle{On emerging entity detection}. In
  \bibinfo{booktitle}{\emph{European Knowledge Acquisition Workshop}}.
  Springer, \bibinfo{pages}{223--238}.
\newblock


\bibitem[\protect\citeauthoryear{Fetahu, Anand, and Koutraki}{Fetahu
  et~al\mbox{.}}{2019}]%
        {fetahu2019tablenet}
\bibfield{author}{\bibinfo{person}{Besnik Fetahu}, \bibinfo{person}{Avishek
  Anand}, {and} \bibinfo{person}{Maria Koutraki}.}
  \bibinfo{year}{2019}\natexlab{}.
\newblock \showarticletitle{Table{N}et: An approach for determining
  fine-grained relations for {W}ikipedia tables}. In
  \bibinfo{booktitle}{\emph{The World Wide Web Conference}}.
  \bibinfo{pages}{2736--2742}.
\newblock


\bibitem[\protect\citeauthoryear{Gal{\'a}rraga, Teflioudi, Hose, and
  Suchanek}{Gal{\'a}rraga et~al\mbox{.}}{2015}]%
        {galarraga2015fast}
\bibfield{author}{\bibinfo{person}{Luis Gal{\'a}rraga},
  \bibinfo{person}{Christina Teflioudi}, \bibinfo{person}{Katja Hose}, {and}
  \bibinfo{person}{Fabian~M Suchanek}.} \bibinfo{year}{2015}\natexlab{}.
\newblock \showarticletitle{Fast rule mining in ontological knowledge bases
  with AMIE+}.
\newblock \bibinfo{journal}{\emph{The VLDB Journal}} \bibinfo{volume}{24},
  \bibinfo{number}{6} (\bibinfo{year}{2015}), \bibinfo{pages}{707--730}.
\newblock


\bibitem[\protect\citeauthoryear{Heist, Hertling, Ringler, and Paulheim}{Heist
  et~al\mbox{.}}{2020}]%
        {heist2020knowledge}
\bibfield{author}{\bibinfo{person}{Nicolas Heist}, \bibinfo{person}{Sven
  Hertling}, \bibinfo{person}{Daniel Ringler}, {and} \bibinfo{person}{Heiko
  Paulheim}.} \bibinfo{year}{2020}\natexlab{}.
\newblock \showarticletitle{Knowledge Graphs on the Web--an Overview}.
\newblock \bibinfo{journal}{\emph{Studies on the Semantic Web}}
  \bibinfo{volume}{47} (\bibinfo{year}{2020}), \bibinfo{pages}{3--22}.
\newblock


\bibitem[\protect\citeauthoryear{Heist and Paulheim}{Heist and
  Paulheim}{2019}]%
        {heist2019uncovering}
\bibfield{author}{\bibinfo{person}{Nicolas Heist} {and} \bibinfo{person}{Heiko
  Paulheim}.} \bibinfo{year}{2019}\natexlab{}.
\newblock \showarticletitle{Uncovering the Semantics of {W}ikipedia
  Categories}. In \bibinfo{booktitle}{\emph{International Semantic Web
  Conference}}. Springer, \bibinfo{pages}{219--236}.
\newblock


\bibitem[\protect\citeauthoryear{Heist and Paulheim}{Heist and
  Paulheim}{2020}]%
        {heist2020entity}
\bibfield{author}{\bibinfo{person}{Nicolas Heist} {and} \bibinfo{person}{Heiko
  Paulheim}.} \bibinfo{year}{2020}\natexlab{}.
\newblock \showarticletitle{Entity Extraction from {Wikipedia} List Pages}. In
  \bibinfo{booktitle}{\emph{Extended Semantic Web Conference}}. Springer,
  \bibinfo{pages}{327--342}.
\newblock


\bibitem[\protect\citeauthoryear{Hertling and Paulheim}{Hertling and
  Paulheim}{2020}]%
        {hertling2020dbkwik}
\bibfield{author}{\bibinfo{person}{Sven Hertling} {and} \bibinfo{person}{Heiko
  Paulheim}.} \bibinfo{year}{2020}\natexlab{}.
\newblock \showarticletitle{{DBkWik}: extracting and integrating knowledge from
  thousands of wikis}.
\newblock \bibinfo{journal}{\emph{Knowledge and Information Systems}}
  \bibinfo{volume}{62}, \bibinfo{number}{6} (\bibinfo{year}{2020}),
  \bibinfo{pages}{2169--2190}.
\newblock


\bibitem[\protect\citeauthoryear{Lehmann}{Lehmann}{2009}]%
        {lehmann2009dl}
\bibfield{author}{\bibinfo{person}{Jens Lehmann}.}
  \bibinfo{year}{2009}\natexlab{}.
\newblock \showarticletitle{DL-Learner: learning concepts in description
  logics}.
\newblock \bibinfo{journal}{\emph{The Journal of Machine Learning Research}}
  \bibinfo{volume}{10} (\bibinfo{year}{2009}), \bibinfo{pages}{2639--2642}.
\newblock


\bibitem[\protect\citeauthoryear{Lehmann et~al\mbox{.}}{Lehmann
  et~al\mbox{.}}{2015}]%
        {lehmann2015dbpedia}
\bibfield{author}{\bibinfo{person}{Jens Lehmann} {et~al\mbox{.}}}
  \bibinfo{year}{2015}\natexlab{}.
\newblock \showarticletitle{DBpedia--a large-scale, multilingual knowledge base
  extracted from Wikipedia}.
\newblock \bibinfo{journal}{\emph{Semantic Web}} \bibinfo{volume}{6},
  \bibinfo{number}{2} (\bibinfo{year}{2015}), \bibinfo{pages}{167--195}.
\newblock


\bibitem[\protect\citeauthoryear{Lehmberg and Bizer}{Lehmberg and
  Bizer}{2017}]%
        {lehmberg2017stitching}
\bibfield{author}{\bibinfo{person}{Oliver Lehmberg} {and}
  \bibinfo{person}{Christian Bizer}.} \bibinfo{year}{2017}\natexlab{}.
\newblock \showarticletitle{Stitching web tables for improving matching
  quality}.
\newblock \bibinfo{journal}{\emph{VLDB Endowment}} \bibinfo{volume}{10},
  \bibinfo{number}{11} (\bibinfo{year}{2017}), \bibinfo{pages}{1502--1513}.
\newblock


\bibitem[\protect\citeauthoryear{Liu, Li, Wang, Sun, and Li}{Liu
  et~al\mbox{.}}{2020}]%
        {liu2020extracting}
\bibfield{author}{\bibinfo{person}{Guiliang Liu}, \bibinfo{person}{Xu Li},
  \bibinfo{person}{Jiakang Wang}, \bibinfo{person}{Mingming Sun}, {and}
  \bibinfo{person}{Ping Li}.} \bibinfo{year}{2020}\natexlab{}.
\newblock \showarticletitle{Extracting Knowledge from Web Text with Monte Carlo
  Tree Search}. In \bibinfo{booktitle}{\emph{The Web Conference 2020}}.
  \bibinfo{pages}{2585--2591}.
\newblock


\bibitem[\protect\citeauthoryear{Macdonald and Barbosa}{Macdonald and
  Barbosa}{2020}]%
        {macdonald2020neural}
\bibfield{author}{\bibinfo{person}{Erin Macdonald} {and}
  \bibinfo{person}{Denilson Barbosa}.} \bibinfo{year}{2020}\natexlab{}.
\newblock \showarticletitle{Neural Relation Extraction on Wikipedia Tables for
  Augmenting Knowledge Graphs}. In \bibinfo{booktitle}{\emph{29th ACM
  International Conference on Information \& Knowledge Management}}.
  \bibinfo{pages}{2133--2136}.
\newblock


\bibitem[\protect\citeauthoryear{Meilicke, Chekol, Ruffinelli, and
  Stuckenschmidt}{Meilicke et~al\mbox{.}}{2019}]%
        {meilicke2019anytime}
\bibfield{author}{\bibinfo{person}{Christian Meilicke},
  \bibinfo{person}{Melisachew~Wudage Chekol}, \bibinfo{person}{Daniel
  Ruffinelli}, {and} \bibinfo{person}{Heiner Stuckenschmidt}.}
  \bibinfo{year}{2019}\natexlab{}.
\newblock \showarticletitle{Anytime Bottom-Up Rule Learning for Knowledge Graph
  Completion.}. In \bibinfo{booktitle}{\emph{28th International Joint
  Conference on Artificial Intelligence}}. \bibinfo{pages}{3137--3143}.
\newblock


\bibitem[\protect\citeauthoryear{Mendes et~al\mbox{.}}{Mendes
  et~al\mbox{.}}{2011}]%
        {mendes2011dbpedia}
\bibfield{author}{\bibinfo{person}{Pablo~N Mendes} {et~al\mbox{.}}}
  \bibinfo{year}{2011}\natexlab{}.
\newblock \showarticletitle{{DB}pedia spotlight: shedding light on the web of
  documents}. In \bibinfo{booktitle}{\emph{7th international conference on
  semantic systems}}. \bibinfo{pages}{1--8}.
\newblock


\bibitem[\protect\citeauthoryear{Meusel, Petrovski, and Bizer}{Meusel
  et~al\mbox{.}}{2014}]%
        {meusel2014webdatacommons}
\bibfield{author}{\bibinfo{person}{Robert Meusel}, \bibinfo{person}{Petar
  Petrovski}, {and} \bibinfo{person}{Christian Bizer}.}
  \bibinfo{year}{2014}\natexlab{}.
\newblock \showarticletitle{The webdatacommons microdata, rdfa and microformat
  dataset series}. In \bibinfo{booktitle}{\emph{International Semantic Web
  Conference}}. Springer, \bibinfo{pages}{277--292}.
\newblock


\bibitem[\protect\citeauthoryear{Oulabi and Bizer}{Oulabi and Bizer}{2019}]%
        {oulabi2019using}
\bibfield{author}{\bibinfo{person}{Yaser Oulabi} {and}
  \bibinfo{person}{Christian Bizer}.} \bibinfo{year}{2019}\natexlab{}.
\newblock \showarticletitle{Using weak supervision to identify long-tail
  entities for knowledge base completion}. In
  \bibinfo{booktitle}{\emph{International Conference on Semantic Systems}}.
  Springer, \bibinfo{pages}{83--98}.
\newblock


\bibitem[\protect\citeauthoryear{Paulheim}{Paulheim}{2017}]%
        {paulheim2017knowledge}
\bibfield{author}{\bibinfo{person}{Heiko Paulheim}.}
  \bibinfo{year}{2017}\natexlab{}.
\newblock \showarticletitle{Knowledge graph refinement: A survey of approaches
  and evaluation methods}.
\newblock \bibinfo{journal}{\emph{Semantic web}} \bibinfo{volume}{8},
  \bibinfo{number}{3} (\bibinfo{year}{2017}), \bibinfo{pages}{489--508}.
\newblock


\bibitem[\protect\citeauthoryear{Paulheim and F{\"u}mkranz}{Paulheim and
  F{\"u}mkranz}{2012}]%
        {paulheim2012unsupervised}
\bibfield{author}{\bibinfo{person}{Heiko Paulheim} {and}
  \bibinfo{person}{Johannes F{\"u}mkranz}.} \bibinfo{year}{2012}\natexlab{}.
\newblock \showarticletitle{Unsupervised generation of data mining features
  from linked open data}. In \bibinfo{booktitle}{\emph{2nd international
  conference on web intelligence, mining and semantics}}.
  \bibinfo{pages}{1--12}.
\newblock


\bibitem[\protect\citeauthoryear{Paulheim and Ponzetto}{Paulheim and
  Ponzetto}{2013}]%
        {paulheim2013extending}
\bibfield{author}{\bibinfo{person}{Heiko Paulheim} {and}
  \bibinfo{person}{Simone~Paolo Ponzetto}.} \bibinfo{year}{2013}\natexlab{}.
\newblock \showarticletitle{Extending DBpedia with Wikipedia List Pages}. In
  \bibinfo{booktitle}{\emph{1st International Workshop on NLP and DBpedia}},
  Vol.~\bibinfo{volume}{1064}. \bibinfo{publisher}{CEUR Workshop Proceedings},
  \bibinfo{pages}{85--90}.
\newblock


\bibitem[\protect\citeauthoryear{Ritze, Lehmberg, Oulabi, and Bizer}{Ritze
  et~al\mbox{.}}{2016}]%
        {ritze2016profiling}
\bibfield{author}{\bibinfo{person}{Dominique Ritze}, \bibinfo{person}{Oliver
  Lehmberg}, \bibinfo{person}{Yaser Oulabi}, {and} \bibinfo{person}{Christian
  Bizer}.} \bibinfo{year}{2016}\natexlab{}.
\newblock \showarticletitle{Profiling the potential of web tables for
  augmenting cross-domain knowledge bases}. In \bibinfo{booktitle}{\emph{The
  World Wide Web Conference}}. \bibinfo{pages}{251--261}.
\newblock


\bibitem[\protect\citeauthoryear{Sakor, Singh, Patel, and Vidal}{Sakor
  et~al\mbox{.}}{2020}]%
        {sakor2020falcon}
\bibfield{author}{\bibinfo{person}{Ahmad Sakor}, \bibinfo{person}{Kuldeep
  Singh}, \bibinfo{person}{Anery Patel}, {and} \bibinfo{person}{Maria-Esther
  Vidal}.} \bibinfo{year}{2020}\natexlab{}.
\newblock \showarticletitle{Falcon 2.0: An entity and relation linking tool
  over {W}ikidata}. In \bibinfo{booktitle}{\emph{29th ACM International
  Conference on Information \& Knowledge Management}}.
  \bibinfo{pages}{3141--3148}.
\newblock


\bibitem[\protect\citeauthoryear{Stanovsky, Michael, Zettlemoyer, and
  Dagan}{Stanovsky et~al\mbox{.}}{2018}]%
        {stanovsky2018supervised}
\bibfield{author}{\bibinfo{person}{Gabriel Stanovsky}, \bibinfo{person}{Julian
  Michael}, \bibinfo{person}{Luke Zettlemoyer}, {and} \bibinfo{person}{Ido
  Dagan}.} \bibinfo{year}{2018}\natexlab{}.
\newblock \showarticletitle{Supervised open information extraction}. In
  \bibinfo{booktitle}{\emph{2018 Conference of the North American Chapter of
  the Association for Computational Linguistics: Human Language Technologies,
  Volume 1 (Long Papers)}}. \bibinfo{pages}{885--895}.
\newblock


\bibitem[\protect\citeauthoryear{Suchanek, Kasneci, and Weikum}{Suchanek
  et~al\mbox{.}}{2007}]%
        {suchanek2007yago}
\bibfield{author}{\bibinfo{person}{Fabian~M Suchanek}, \bibinfo{person}{Gjergji
  Kasneci}, {and} \bibinfo{person}{Gerhard Weikum}.}
  \bibinfo{year}{2007}\natexlab{}.
\newblock \showarticletitle{Yago: a core of semantic knowledge}. In
  \bibinfo{booktitle}{\emph{The World Wide Web Conference}}.
  \bibinfo{pages}{697--706}.
\newblock


\bibitem[\protect\citeauthoryear{Van~Erp, Mendes, Paulheim, Ilievski, Plu,
  Rizzo, and Waitelonis}{Van~Erp et~al\mbox{.}}{2016}]%
        {van2016evaluating}
\bibfield{author}{\bibinfo{person}{Marieke Van~Erp}, \bibinfo{person}{Pablo
  Mendes}, \bibinfo{person}{Heiko Paulheim}, \bibinfo{person}{Filip Ilievski},
  \bibinfo{person}{Julien Plu}, \bibinfo{person}{Giuseppe Rizzo}, {and}
  \bibinfo{person}{J{\"o}rg Waitelonis}.} \bibinfo{year}{2016}\natexlab{}.
\newblock \showarticletitle{Evaluating entity linking: An analysis of current
  benchmark datasets and a roadmap for doing a better job}. In
  \bibinfo{booktitle}{\emph{Proceedings of the Tenth International Conference
  on Language Resources and Evaluation (LREC'16)}}.
  \bibinfo{pages}{4373--4379}.
\newblock


\bibitem[\protect\citeauthoryear{V{\"o}lker and Niepert}{V{\"o}lker and
  Niepert}{2011}]%
        {volker2011statistical}
\bibfield{author}{\bibinfo{person}{Johanna V{\"o}lker} {and}
  \bibinfo{person}{Mathias Niepert}.} \bibinfo{year}{2011}\natexlab{}.
\newblock \showarticletitle{Statistical schema induction}. In
  \bibinfo{booktitle}{\emph{Extended Semantic Web Conference}}. Springer,
  \bibinfo{pages}{124--138}.
\newblock


\bibitem[\protect\citeauthoryear{Vrande{\v{c}}i{\'c} and
  Kr{\"o}tzsch}{Vrande{\v{c}}i{\'c} and Kr{\"o}tzsch}{2014}]%
        {vrandevcic2014wikidata}
\bibfield{author}{\bibinfo{person}{Denny Vrande{\v{c}}i{\'c}} {and}
  \bibinfo{person}{Markus Kr{\"o}tzsch}.} \bibinfo{year}{2014}\natexlab{}.
\newblock \showarticletitle{Wikidata: a free collaborative knowledgebase}.
\newblock \bibinfo{journal}{\emph{Commun. ACM}} \bibinfo{volume}{57},
  \bibinfo{number}{10} (\bibinfo{year}{2014}), \bibinfo{pages}{78--85}.
\newblock


\bibitem[\protect\citeauthoryear{Wang, Mao, Wang, and Guo}{Wang
  et~al\mbox{.}}{2017}]%
        {wang2017knowledge}
\bibfield{author}{\bibinfo{person}{Quan Wang}, \bibinfo{person}{Zhendong Mao},
  \bibinfo{person}{Bin Wang}, {and} \bibinfo{person}{Li Guo}.}
  \bibinfo{year}{2017}\natexlab{}.
\newblock \showarticletitle{Knowledge graph embedding: A survey of approaches
  and applications}.
\newblock \bibinfo{journal}{\emph{IEEE Transactions on Knowledge and Data
  Engineering}} \bibinfo{volume}{29}, \bibinfo{number}{12}
  (\bibinfo{year}{2017}), \bibinfo{pages}{2724--2743}.
\newblock


\bibitem[\protect\citeauthoryear{Xu, Xie, Zhang, Xiao, Wang, and Wang}{Xu
  et~al\mbox{.}}{2016}]%
        {xu2016learning}
\bibfield{author}{\bibinfo{person}{Bo Xu}, \bibinfo{person}{Chenhao Xie},
  \bibinfo{person}{Yi Zhang}, \bibinfo{person}{Yanghua Xiao},
  \bibinfo{person}{Haixun Wang}, {and} \bibinfo{person}{Wei Wang}.}
  \bibinfo{year}{2016}\natexlab{}.
\newblock \showarticletitle{Learning defining features for categories}. In
  \bibinfo{booktitle}{\emph{25th International Joint Conference on Artificial
  Intelligence}}. \bibinfo{pages}{3924--3930}.
\newblock


\bibitem[\protect\citeauthoryear{Yakout, Ganjam, Chakrabarti, and
  Chaudhuri}{Yakout et~al\mbox{.}}{2012}]%
        {yakout2012infogather}
\bibfield{author}{\bibinfo{person}{Mohamed Yakout}, \bibinfo{person}{Kris
  Ganjam}, \bibinfo{person}{Kaushik Chakrabarti}, {and}
  \bibinfo{person}{Surajit Chaudhuri}.} \bibinfo{year}{2012}\natexlab{}.
\newblock \showarticletitle{Info{G}ather: entity augmentation and attribute
  discovery by holistic matching with web tables}. In
  \bibinfo{booktitle}{\emph{2012 ACM SIGMOD International Conference on
  Management of Data}}. \bibinfo{pages}{97--108}.
\newblock


\bibitem[\protect\citeauthoryear{Zhang et~al\mbox{.}}{Zhang
  et~al\mbox{.}}{2020b}]%
        {zhang2020novel}
\bibfield{author}{\bibinfo{person}{Shuo Zhang} {et~al\mbox{.}}}
  \bibinfo{year}{2020}\natexlab{b}.
\newblock \showarticletitle{Novel Entity Discovery from Web Tables}. In
  \bibinfo{booktitle}{\emph{The Web Conference 2020}}.
  \bibinfo{pages}{1298--1308}.
\newblock


\bibitem[\protect\citeauthoryear{Zhang and Balog}{Zhang and Balog}{2020}]%
        {zhang2020web}
\bibfield{author}{\bibinfo{person}{Shuo Zhang} {and} \bibinfo{person}{Krisztian
  Balog}.} \bibinfo{year}{2020}\natexlab{}.
\newblock \showarticletitle{Web Table Extraction, Retrieval, and Augmentation:
  A Survey}.
\newblock \bibinfo{journal}{\emph{ACM Transactions on Intelligent Systems and
  Technology (TIST)}} \bibinfo{volume}{11}, \bibinfo{number}{2}
  (\bibinfo{year}{2020}), \bibinfo{pages}{1--35}.
\newblock


\bibitem[\protect\citeauthoryear{Zhang, Balog, and Callan}{Zhang
  et~al\mbox{.}}{2020a}]%
        {zhang2020generating}
\bibfield{author}{\bibinfo{person}{Shuo Zhang}, \bibinfo{person}{Krisztian
  Balog}, {and} \bibinfo{person}{Jamie Callan}.}
  \bibinfo{year}{2020}\natexlab{a}.
\newblock \showarticletitle{Generating Categories for Sets of Entities}. In
  \bibinfo{booktitle}{\emph{29th ACM International Conference on Information \&
  Knowledge Management}}. \bibinfo{pages}{1833--1842}.
\newblock


\bibitem[\protect\citeauthoryear{Zhang}{Zhang}{2014}]%
        {zhang2014towards}
\bibfield{author}{\bibinfo{person}{Ziqi Zhang}.}
  \bibinfo{year}{2014}\natexlab{}.
\newblock \showarticletitle{Towards efficient and effective semantic table
  interpretation}. In \bibinfo{booktitle}{\emph{International Semantic Web
  Conference}}. Springer, \bibinfo{pages}{487--502}.
\newblock


\bibitem[\protect\citeauthoryear{Zhu and Iglesias}{Zhu and Iglesias}{2018}]%
        {zhu2018exploiting}
\bibfield{author}{\bibinfo{person}{Ganggao Zhu} {and} \bibinfo{person}{Carlos~A
  Iglesias}.} \bibinfo{year}{2018}\natexlab{}.
\newblock \showarticletitle{Exploiting semantic similarity for named entity
  disambiguation in knowledge graphs}.
\newblock \bibinfo{journal}{\emph{Expert Systems with Applications}}
  \bibinfo{volume}{101} (\bibinfo{year}{2018}), \bibinfo{pages}{8--24}.
\newblock


\end{thebibliography}
\end{document}